\documentclass[aps,prb, twocolumn,groupedaddress,floatfix,showpacs]{revtex4}

\usepackage{graphicx}
\usepackage{graphics}
\usepackage{amsmath} 
\usepackage{ulem}
\newcommand{\grad}{\ensuremath{^\circ}}

\begin{document}


\title{Green Function theory vs. Quantum Monte Carlo Calculation for thin magnetic films}



\author{S. Henning}
\email{henning@physik.hu-berlin.de}
\author{F. K\"{o}rmann}
\author{J. Kienert}
\author{W. Nolting}\affiliation{Lehrstuhl Festk\"{o}rpertheorie, Institut f\"ur Physik, Humboldt-Universit\"{a}t zu Berlin, Newtonstrasse 15, 12489 Berlin, Germany}
\author{S. Schwieger}\affiliation{Technische Universit\"at Ilmenau, Theoretische Physik I, Postfach 10 05 65, 98684 Ilmenau, Germany}


\date{\today}

\begin{abstract}
In this work we compare numerically exact Quantum Monte Carlo (QMC) calculations and 
Green function theory (GFT) calculations of thin ferromagnetic films including second order anisotropies.
Thereby we concentrate on easy plane systems, i.e. systems for which the anisotropy 
favors a magnetization parallel to the film plane. We discuss these 
systems in perpendicular external field, i.e. $B$ parallel to the film normal. GFT results are 
in good agreement with QMC for high enough fields and temperatures. Below a critical field 
or a critical temperature no collinear stable magnetization exists in GFT. On the other hand 
QMC gives finite magnetization even below those critical values. This indicates that there 
occurs a transition from non-collinear to collinear configurations with increasing field or 
temperature. For slightly tilted external fields a rotation of magnetization from out-of-plane
to in-plane orientation is found with decreasing temperature. 
\end{abstract}


\maketitle

\section{Introduction}
The fast development of technological applications based on magnetic systems in the last years, 
e.g. magnetic data storage devices, causes a high interest in thin magnetic films. One 
precondition for the technological development is the investigation of magnetic anisotropies 
and spin reorientation transitions connected therewith. Those reorientation transitions can 
occur from out-of-plane to in-plane or vice versa for increasing film thickness $d$\cite{farle},
temperature 
$T$ \cite{wandlitz17,wandlitz18,wandlitz19,wandlitz20,wandlitz21,wandlitz22,wandlitz24}, or 
external field $B_0$.\\ 
\indent Quantum Monte Carlo (QMC) calculations give the possibility to compare numerically 
exact results with 
analytical approximations.
In Ref.~\onlinecite{qm} the authors investigated a ferromagnetic monolayer including positive 
second order anisotropy (easy axis perpendicular to the film plane). They discuss the 
temperature dependence of the magnetization $\langle S_{z}\rangle(T)$ as well as field induced 
reorientation transitions from out-of-plane to in-plane and compare the QMC results with Green 
function theory (GFT).
They found good agreement in the case of applied external field in the easy direction 
(here $z$-axis). However, their GFT fails for external field applied in arbitrary 
direction, especially in the hard direction (within the film plane). As shown in 
Ref.~\onlinecite{schwieger1} for getting closer to the QMC results for magnetic field induced 
reorientation from out-of-plane to in-plane a more careful treatment of the local anisotropy 
terms is needed. In Refs.~\onlinecite{schwieger1,schwieger2,dipolpaper,pini} a 
decoupling scheme was presented which yields excellent agreement with QMC results for 
out-of-plane systems.\\ 
\indent The availability of theories such as GFT and their check against state-of-the-art 
numerical algorithms is highly desirable because of the size limitations of systems where
QMC can be performed. On the other hand the extension of GFT from a monolayer (where it can be 
compared to QMC as in the the present work) to multilayer systems is a straightforward
task without further approximations\cite{schwieger2}.\\ 
\indent Up to now, to our knowledge, there is no comparison between QMC and 
approximative theories for easy-plane systems and it is not obvious  that the theory 
presented in Refs.~\onlinecite{schwieger1,schwieger2,dipolpaper,pini} can reproduce the QMC results for in-plane 
systems as accurately as for the out-of-plane case. In contrast to the easy-axis case 
where a certain direction is preferred
by the single ion anisotropy in easy-plane systems the full xy-plane is favored and no
particular direction is distinguished within the plane. 
A magnetic field applied perpendicular to the plane does not destroy the $xy$-symmetry.\\ 
\indent For systems exhibiting this kind of symmetry it was shown in a classical 
treatment that for external fields
smaller than a critical field $0\le B<B_{crit}$ ($B\:||\:z$) stable vortices, i.e. a non-collinear 
arrangement of spins, can exist\cite{wysin98, vedmedenko99, lee04, rapini07, ivanov02}. 
These vortices can
undergo a Berezinskii-Kosterlitz-Thouless  (BKT) transition\cite{kosterlitz73}. Depending on 
the strength of the anisotropy $K_2$ there might be vortices with or without
a finite z-component of magnetization\cite{wysin98}. In the small anisotropy case 
(which is considered in this work, $|K_2|< 0.1J$) there is a finite out-of-plane component and
for zero field the two possible directions of magnetization ($\pm z$) are 
energetically degenerate. For increasing magnetic field in z-direction the
vortices antiparallel to the field become more and more unstable (heavy vortices). 
However the so called light vortices (parallel to the field) are stable up to a 
critical field $B_z=B_{crit}$ and contribute a finite $z$-component to the net-magnetization 
of the considered system \cite{ivanov02}.\\
\indent The vortices in connection with a finite z-component of the net-magnetization emerge 
because of two reasons: first the 
competition between the anisotropy (favoring a orientation of the magnetization within the 
$xy$-plane) and the external field (favoring a perpendicular magnetization), and second: 
the $xy$-symmetry of the system, which does not allow for a rotated homogeneous phase.\\
\indent In this paper we investigate both aspects, i.e. the field vs. anisotropy competition
as well as the symmetry properties in detail for a \textit{quantum mechanical} system. We will 
compare the results of QMC and GFT calculations.\\ 
\indent As explained in more detail below, the QMC algorithm used here
allows only for an external field applied in $z$-direction. Thus the $xy$-symmetry can not be
broken and no comparison between $xy$-symmetric and asymmetric systems is possible. We will 
use GFT to clarify the influence of this symmetry breaking on the homogeneous phase. On the
other hand, the GFT used here is by ansatz limited to the homogeneous phase. Therefore it 
can not describe a non-collinear (e.g, vortex-) magnetic phase, which is expected 
for $B\:||\:z$ and small field strengths. The breakdown of magnetization in GFT as well as
an exposed maximum in the magnetization in QMC at certain critical values of the external
field or temperature gives, however, a clear
fingerprint of non-collinear configurations, at least if there is no meta-stable homogeneous
phase. Below these critical values there will be a finite $z$-component in QMC 
\textit{and} a vanishing magnetization in GFT.\\ 
For parameters, where both theories are applicable, 
QMC serves as a test for the approximations needed in GFT.\\\\
\indent In this work we find indications for non-collinear spin configurations below a critical
field or temperature for $B\:||\:z$ by comparing results of QMC and GFT as explained in the last
clause.
Above the critical field we obtain good agreement between QMC and GFT results. Breaking the
$xy$-symmetry by adding a small $x$-component to the external field  yields a stable collinear
solution in GFT. The $z$-component of the magnetization in this case is in good agreement with the QMC 
results calculated with untilted field. Thus we can conclude that except for the restriction to
collinear magnetic states GFT describes the competition between external field and anisotropy
quite well.\\\\ 
\indent The paper is organized as follows: First we explain the basics of the GFT and the 
QMC calculations. Then we apply both approaches to easy-plane systems in external magnetic fields 
and report the results of our calculations. 

\section{Theory}

\subsection{Green Function Theory}\label{GF}
In the following we present our theoretical approach using Green function theory.
The focus of this work lies on the translational invariant system of a two-dimensional 
monolayer. 
Therefore the following Hamiltonian is used: 
\begin{eqnarray} H &=& -\frac{1}{2}\sum_{ij}J_{ij}\mathbf{S}_{i}\mathbf{S}_{j} 
               - \mathbf{B}\sum_{i}\mathbf{S}_{i} - K_2\sum_{i}(S_{z,i})^2. \label{hamilt}
 \end{eqnarray} 
The first term describes the Heisenberg coupling $J_{ij}$ between spins
$\mathbf{S}_{i}$ and $\mathbf{S}_{j}$ located at sites $i$ and $j$. 
The second term contains an external magnetic field $\mathbf{B}$ in arbitrary direction 
(the Land{$\rm\acute e$} factor $g_{J}$ and the Bohr magneton $\mu_{B}$ are absorbed in 
$\mathbf{B}$).  The third term represents second order lattice anisotropy due to spin-orbit coupling. $S_{z,i}$ 
is the $z$-component of $\mathbf{S}_{i}$ (the $z$-axis of the coordinate system is oriented 
perpendicular to the film-plane).
The lattice anisotropy favours in-plane ($K_2<0$) or out-of-plane ($K_2>0$) orientation.
Our Hamiltonian is similar to that used in 
Refs.~\onlinecite{wandlitz,jensen2000,schwieger1,schwieger2,pini} 
for the investigation of the magnetic anisotropy and the field induced reorientation transition.
To simplify calculations we consider nearest neighbor coupling only 
\begin{eqnarray} 
J_{ij} = \left\{ \begin{array}{cc} J & ~~(i),(j) \text{ n.n.}\\ 0 & \text{otherwise.} 
\end{array}\right. 
\end{eqnarray} 
The main idea of the special treatment presented in Refs.~\onlinecite{schwieger1,schwieger2,dipolpaper,pini} 
is that, before 
any decoupling is applied, the coordinate system $\Sigma$ is rotated to a new system 
$\Sigma^{\prime}$ where the new $z^{\prime}$-axis is parallel to the magnetization
implying a collinear alignment of all spins within the layer.
Then a combination of Random Phase approximation (RPA)\cite{RPA}  for the nonlocal terms in Eq. 
(\ref{hamilt}) (Heisenberg exchange interaction term) and Anderson-Callen 
approximation (AC)\cite{ac} for the local lattice anisotropy term is applied in the rotated 
system.  
After application of the approximation one gets an \textit{effective} anisotropy 
\begin{eqnarray}
K_{eff}(T)=2K_{2} \left(1-\frac{1}{2S^2}\left( S(S+1) - \langle S^2_{z\prime}\rangle
\right)\right)\langle S_{z\prime}\rangle\label{Keff}
\end{eqnarray}
where $\langle S_{z\prime}\rangle$ is the norm of the magnetization and $S$ is the 
spin quantum number, that we have chosen to be $S=1$ in all our calculations.\\
\indent As shown in comparison with an exact treatment of the local anisotropy term in 
Ref.~\onlinecite{exakt}
this approximation still holds up to anisotropy strengths $K_2\sim 1/2J$.
Therefore we restrict ourselves in the following to small anisotropies ($K_2\leq 0.1J$) as 
found in most real materials\footnote{Besides some rare earth materials where the anisotropy 
can be of the order of $J$.}.
For a magnetic field applied in the $xz$-plane ($\mathbf{B}=(B_x,0,B_z)$) our theory gives 
a condition for the polar angle $\theta$ of the magnetization:
\begin{equation}
\sin\theta B_{z}-\cos\theta B_{x}+K_{eff}\sin\theta\cos\theta=0
\end{equation}
The uniform magnon energies ($\mathbf{q}=\mathbf{0}$) which dominate the physical 
behavior of the magnetic system can easily be extracted from the theory \cite{dipolpaper,pini}:
\begin{multline}
E_{\mathbf{q}=\mathbf{0}}^2=\left(\cos\theta B_z +\sin\theta B_x+K_{eff}(\cos^2\theta-\sin^2\theta)\right)\cdot\\
\left(\cos\theta B_z+\sin\theta B_x+K_{eff}\cos^2\theta\right)
\label{uniform_modes}
\end{multline}
This result 
coincides with the spin-wave result \cite{pini} if one replaces $\langle S_{z\prime}\rangle$ 
by the spin quantum number $S$ and $K_{eff}$ by the bare anisotropy constant $K_2$ in 
Eq.~(\ref{uniform_modes}). 
For an easy-plane system ($K_{eff}<0$) with external field $B$ in $z$-direction the 
polar angle 
$\theta$ of the magnetization\footnote{For $B<|K_{eff}|$ there is another mathematical 
solution ($\sin\theta=0$) which however is unstable (see appendix A).} is given by:
\begin{eqnarray} 
\cos\theta = \left\{ \begin{array}{cc} -B/K_{eff}(T) & ~~\text{for}~~B<|K_{eff}(T)|\\
 1 & \text{otherwise} 
\end{array}\right. \label{2winkelloesungen}
\end{eqnarray} 
By inserting Eq.(\ref{2winkelloesungen}) into Eq.(\ref{uniform_modes}) one immediately gets: 
\begin{eqnarray} 
E^{K_{eff}<0}_{\mathbf{q}=\mathbf{0}}(B) = \left\{ \begin{array}{cc} 0 
 & B<|K_{eff}(T)|\\
 B+K_{eff}(T) & \text{otherwise.} 
\end{array}\right.\label{EkeffBz}
\end{eqnarray} 
For gapless magnon energies $E_{\mathbf{q}=\mathbf{0}}=0$ the magnon occupation number $\phi$ diverges 
($\phi\rightarrow\infty$) in film systems with ferromagnetic coupling $J>0$ and 
the magnetization becomes 
zero $\langle S_{z\prime}\rangle=0$ in the collinear phase. This
can be seen by following an argument of Bloch\cite{bloch30} already given in 1930. 
Since the 
spin wave dispersion is $E\approx q^2$ in the vicinity of $\mathbf{q}=\mathbf{0}$ the spin-wave
 density of states $N(E)$ is independent of $E$ for a two-dimensional system for $E$ close to
zero. The excitation of spin-waves at finite temperature leads to a variation of the 
magnetization of the order:
\begin{eqnarray}
  \label{eq:magnet_var}
  \Delta m(T) & \sim & \int_{0}^{\infty}\frac{N(E)dE}{\exp (E/k_{B}T)-1)} \nonumber \\
              & \sim & k_BT\int_{0}^{\infty}\frac{dx}{\exp (x)-1}.  
\end{eqnarray}
Since the integral in Eq.~(\ref{eq:magnet_var}) diverges for $T\neq0$ and exited spin-waves 
lead to a reduction of
the magnetization one can conclude that the magnetization should be zero at finite 
temperature.   
However for an infinitesimally small contribution of the external field parallel to the 
plane, i.e. 
$B_x\neq 0$, a finite gap in the excitation spectrum at $\mathbf{q}=\mathbf{0}$ opens.
\begin{figure}
\resizebox{0.9\columnwidth}{!}{
\includegraphics{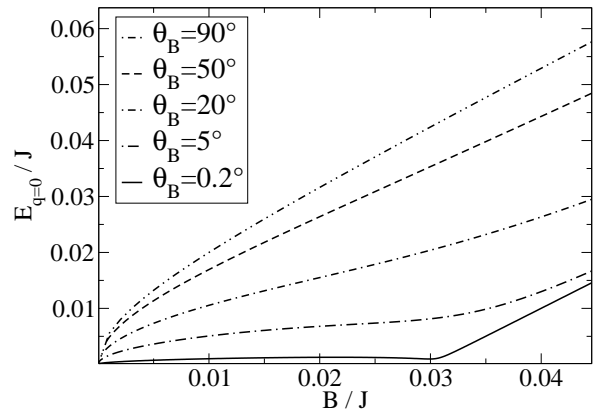}
}
\caption{The energies of the uniform magnon mode $E_{\mathbf{q}=\mathbf{0}}(B)$ 
for different polar angles $\theta_B$ of the external field. $E_{\mathbf{q}=\mathbf{0}}$ is zero below
$B/J \approx 0.03$ for $\theta_B = 0\grad$. The prefactors $g_J \mu_B$ and $k_B$ are absorbed in $B$ and $T$ 
respectively. The latter are given in units of the nearest neighbor Heisenberg coupling $J$.
Parameters: $S=1$, $K/J=-0.03$ and $T/J=10^{-4}$. }\label{magnonnegK}
\end{figure}
This can be seen in Fig.\ref{magnonnegK} where the uniform magnon modes 
$E_{\mathbf{q}=\mathbf{0}}(B)$ are shown for
different orientations $\theta_B$, where $\theta_B$ is the polar angle of the external field. 
The integral (\ref{eq:magnet_var}) is now finite and a stable finite magnetization
in the collinear phase having a well defined orientation in the $xz$-plane is possible.\\
\indent Let us now come back to the case where the applied field is aligned in $z$-direction. 
It can be seen from Eq. (\ref{EkeffBz}) that for
external field $B$ ($B\:||\:z$) larger than a critical field $B>B_{crit}$ given by:
\begin{equation}
B_{crit}\stackrel{!}{=}|K_{eff}(T,B)|\label{criticfield}
\end{equation}
a stable collinear solution exists. 
\begin{figure}
\resizebox{0.9\columnwidth}{!}{
\includegraphics{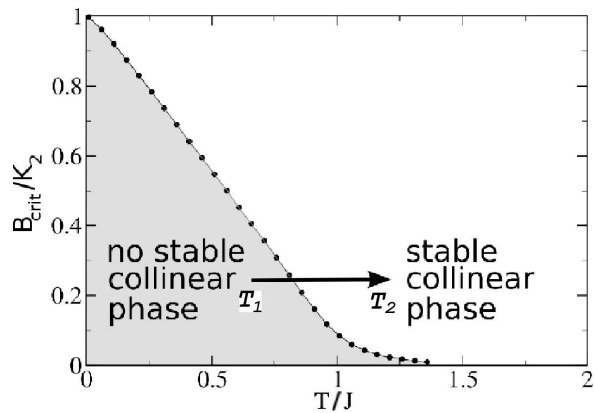}
}
\caption{The normalized critical field $B_{crit}/K_2$ as a function of temperature.
Parameters: $S=1$.}\label{bkrit4}
\end{figure}
Since $K_{eff}(T)$ is a decreasing function of temperature $T$ a transition from non-collinear 
to collinear phase with increasing temperature is possible.
In Fig. \ref{bkrit4} we show the normalized critical field (\ref{criticfield}) $B_{crit}/K_2$ as a 
function of temperature $T$. For a constant magnetic field $B$ ($B\:||\:z$) at
a temperature $T_1$ with $B<K_{eff}(T_1,B)$ no stable collinear phase exist. Then by 
increasing the
temperature up to $T_2$ the effective anisotropy $K_{eff}$ is sufficiently reduced such 
that $B>K_{eff}(T_2,B)$, and the collinear phase becomes stable.
Before we come to the results let us briefly sketch the main aspects of the QMC.

\subsection{QMC}
In the last section we gave a short description of the theory used to treat a system described
by a Hamiltonian of form (\ref{hamilt}). This theory applies to the thermodynamic limit (films
of infinte size) but 
contains certain approximations. Additionally the GFT is restricted to ordered phases with a 
collinear alignment of all spins.
Therefore it would be very useful to have exact results at 
hand to crosscheck the predictions of GFT. A Quantum Monte Carlo method, particularly 
well suited for spin systems, is the stochastic series expansion (SSE) with directed loop 
update. We will sketch this method here only briefly as detailed descriptions can be already 
found elsewhere \cite{sandvik99,sandvik02,alet05}.

Our starting point is the series expansion of the partition function
\begin{equation}
   \label{partition}
   \mathrm{Z} = \mathrm{Tr}e^{-\beta H} = \sum_{n=0}^{\infty}\sum_{\alpha}\frac{\beta^n}{n!}
   \langle\alpha|(-H)^n|\alpha\rangle
\end{equation}
where $H$ denotes the Hamiltonian, $\{|\alpha\rangle\}$ are basis vectors of a 
proper Hilbert space and $\beta$ is the inverse temperature. The Hamiltonian is then rewritten 
in terms of bond Hamiltonians:
\begin{equation}
  \label{bondh}
  H=-J\sum_{b=1}^M H_b
\end{equation}
where $H_b$ can be further decomposed into a diagonal and an off-diagonal part:
\begin{eqnarray}
  \label{hbdecomp}
    H_{D,b} & = & C + S_{i(b)}^{z}S_{j(b)}^{z}+b_b[S_{i(b)}^{z}+S_{j(b)}^{z}]\\
            &   & +k_{2b}[(S_{i(b)}^{z})^2+(S_{j(b)}^{z})^2]\nonumber \\
    H_{O,b} & = & \frac{1}{2}[S_{i(b)}^{+}S_{j(b)}^{-}+S_{i(b)}^{-}S_{j(b)}^{+}]
\end{eqnarray}
Here we have renormalized the anisotropy constant $k_{2b}$ and the magnetic field $b_b$ in such 
a way that (\ref{bondh}) coincides with (\ref{hamilt}). $i(b)$ and $j(b)$ denotes the lattice 
sites connected by the bond $b$ and the additional constant $C$ in $H_{D,b}$ will be chosen 
such that all matrix elements of this term become positive, a condition necessary to 
interpret them as probabilities.
Note that for a finite system at finite temperature the power series of the partition 
function can be truncated at a finite cutoff length $\Lambda$ without introducing any 
systematic error in practical computations\cite{sandvik02}. Therefore reinserting 
(\ref{bondh}) into (\ref{partition}) and rewriting the result yields:
\begin{equation}
  \label{partfin}
  Z = \sum_{n=0}^{\Lambda}\sum_{S_{C_{\Lambda}}}\sum_{\alpha}\frac{\beta^n(\Lambda-n)!}{\Lambda!}
       \langle\alpha|S_{C_{\Lambda}}|\alpha\rangle.
\end{equation}
Here $S_{C_{\Lambda}}$ denotes a product of operators (operator string) consisting of n non-unity operators and 
($\Lambda-n$) unity operators $H_0=\mathrm{Id}$ which were inserted to get operator strings 
of equal length $\Lambda$.\\
In fact it is impossible to evaluate all operator strings in (\ref{partfin}). The SSE-QMC 
replaces such an evaluation therefore by importance sampling over the strings according to 
their relative weight. Hence an efficient scheme for generating new operator strings is needed. 
In the directed loop version of the SSE this is done by dividing the update into two parts. 
In a first step a diagonal update is performed by traversing the operator string and replacing 
some unity operators by diagonal bond operators and vice versa (the probabilities for both 
substitutions have to fulfill the detailed balance criterion). Then the loop update follows 
in which new non-diagonal bond operators can appear in the operator string. For details of 
the update procedure we refer the interested reader to the according literature 
\cite{sandvik99, sandvik02, alet05}.\\
\indent A full implementation of the SSE with directed loop update which we have used for all 
QMC calculations in this work can be found in the ALPS project \cite{ALPS, alet05}.    
Since the SSE-QMC used by us is implemented in z-representation
(spin quantization axis along z-axis) in-plane correlation functions e.g. the in-plane 
magnetization are not accessible. Further $B\:||\:z$ is the only possible field direction
in the used QMC implementation because a traverse field (in-plane field component) would lead to
non-closing loops (see Ref.~\onlinecite{qm}).

\section{Results}

As mentioned in Sec.~\ref{GF} the results for the in-plane systems are very sensitive to 
the effective anisotropy $K_{eff}(T)$. This sensitivity of the anisotropy is less pronounced 
for out-of-plane systems ($K_2>0$) since the applied field $B$ ($B\:||\:z$) and the intrinsic easy 
axis are parallel. In order to test our decoupling scheme (RPA+AC) we first compare GFT and QMC 
for an out-of-plane system.\footnote{Note that a 
similar result has already been published in Ref.~\onlinecite{qm}.}\\
\begin{figure}
\resizebox{0.9\columnwidth}{!}{
\includegraphics{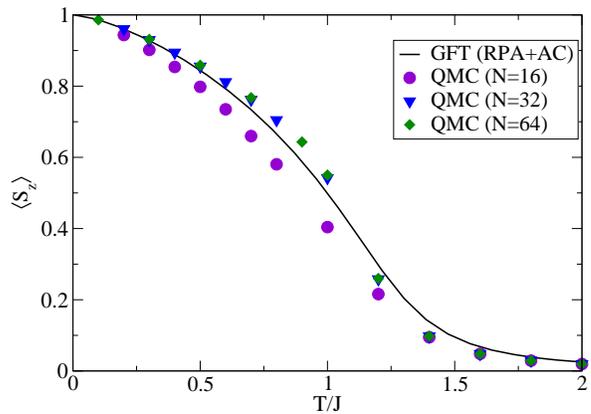}
}
\caption{Magnetization vs. temperature for an out-of-plane easy-axis system ($K_2>0$).
Straight line: GFT (RPA+AC) result; symbols: QMC results for different system sizes $N^2$. 
Parameters: $S=1$, $B/J=0.01$ ($B\:||\:z$) and $K_2/J=0.01$.}\label{figure3}
\end{figure}
\indent In Fig.~\ref{figure3} the magnetization $\langle S_{z}\rangle$ as a function of 
temperature $T$ is shown. The straight line belongs to the GFT whereas the symbols 
show the result of the QMC for different system sizes. 
Let us first comment on finite size effects in the QMC results.
 
It can be seen in Fig.~\ref{figure3}
that the QMC results converge for increasing system size $N^2$ (for $N\times N$ square lattice). 
Indeed for $N\geq 32$ the QMC results are unbiased by finite size effects and resulting 
magnetization curves are almost equal for increasing $N\geq 32$. Note that we have omitted 
error bars in the figures showing QMC results because the relative errors are of the order
$10^{-4}$.

We now compare the GFT with the QMC results ($N=64$). For low 
temperatures ($T/J\leq 0.5$) we obtain excellent quantitative agreement. This is plausible 
because in this region the GFT result coincides with the result of the spin-wave theory which 
is known to be reliable (exact for $T=0$) for low temperatures.
For the intermediate region $T/J=0.5..1$ the RPA slightly underestimates the magnetization 
which was also found in Ref.~\onlinecite{qm}.  
The opposite is the case in the region near the 
extrapolated Curie temperature $T_C$ \footnote{Strictly speaking there is no phase transition
because of the applied magnetic field as can be seen from the large tail of the magnetization
curve. However one can extract a $T_C$ from the curves by extrapolating to the zero field case
and additionally to an infinte system size in the QMC calculations.}, where the magnetization
is overestimated. The reason is the presence of longitudinal fluctuations, which play an important 
role in this region and it is well known that the RPA fails to treat them properly. 
\begin{figure}
\resizebox{0.9\columnwidth}{!}{
\includegraphics{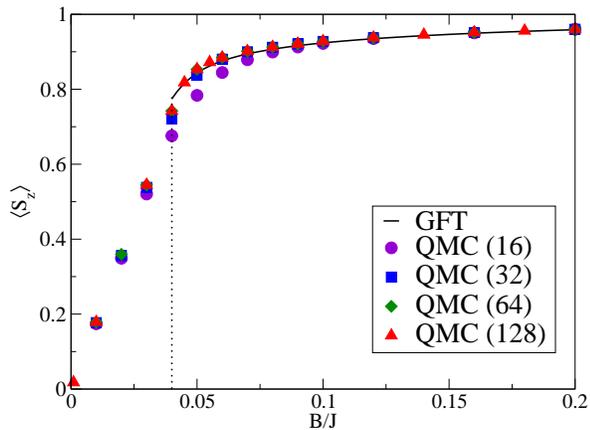}
}
\caption{$z$-component of magnetization as a function of external magnetic field for fixed 
temperature $T/J=0.4$. 
In contrary to the GFT the magnetization obtained by QMC remains finite for all fields.
The QMC results are converged for $N\ge64$. 
Parameters: $S=1$, $K_2/J=-0.06$ and $\theta_B=0\grad$.}\label{figure4}
\end{figure}
\begin{figure}
\resizebox{0.9\columnwidth}{!}{
\includegraphics{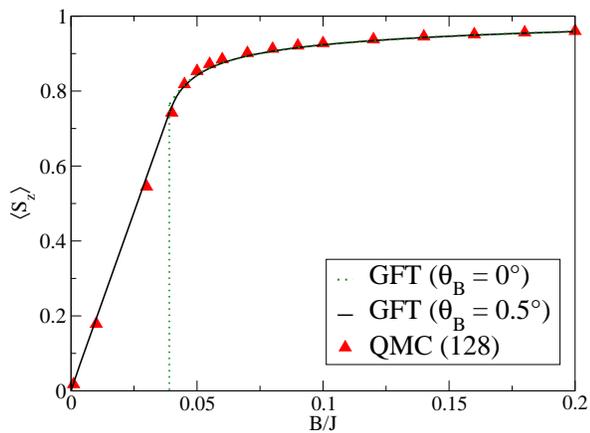}
}
\caption{$z$-component of magnetization vs. external field for $T/J=0.4$ with slightly tilted field 
($\theta_B=0.5\grad$)in the GFT result (solid line). The dotted line shows GFT result for 
($\theta_B=0\grad$). 
Other parameters as in Fig.~\ref{figure4}.}
\label{figure5}
\end{figure}

\begin{figure}
\resizebox{0.9\columnwidth}{!}{
\includegraphics{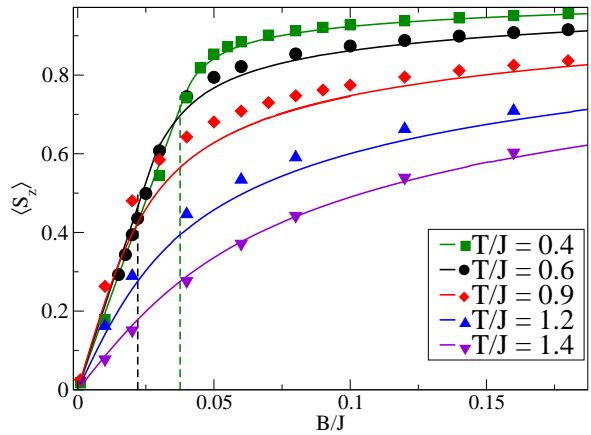}
}
\caption{$z$-component of magnetization vs. external field for different temperatures $T/J$ 
and fixed system size $N^2$ ($N=128$). Solid lines: GFT ($\theta_B=0.5\grad$), dashed lines: GFT ($\theta_B = 0\grad$) 
other parameters as in Fig.~\ref{figure4}.}\label{figure6}
\end{figure}
\indent We consider now the case of in-plane systems ($K_2<0$) and applied field in the 
hard direction ($B\:||\:z$).
As already mentioned there is no 'collinear' magnetization in the 
GFT for $B_z<|K_{eff}(T)|$. 
In Fig.~\ref{figure4} the $z$-component of the magnetization is shown as a
function of the external field $B$ for a constant temperature $T/J=0.4$. 
As in Fig.~\ref{figure3} we see that the QMC results for $N\geq 64$ 
are almost converged and the finite 
size of the calculated system in QMC should not influence the results anymore. 
The dotted line marks a critical field $B_{crit}$. For 
magnetic fields larger than the critical one
$B>B_{crit}$ we obtain good agreement between QMC and GFT results. Below the critical field
$B<B_{crit}$ GFT does not yield a stable homogeneous magnetization. However the QMC
results show that there is a finite $z$-component of the magnetization in the considered system
for $0\leq B \leq B_{crit}$.\\
\indent In order to compare QMC with GFT results we have
tilted the magnetic field by $\theta_B=0.5\grad$ which corresponds to $B_x< 10^{-2}B_z$ in the
GFT. As explained before any symmetry breaking field $B_x\neq 0$ leads to a stable
homogeneous magnetization with well-defined orientation in the $xz$-plane. However such a small
contribution of the external field within the plane should hardly influence the
$z$-component of the magnetization. 
This is confirmed by Fig.~\ref{figure5} where we show QMC results ($N=128$, $\theta_B=0\grad$)
as well as the corresponding GFT results with $\theta_B=0\grad$ and
$\theta_B=0.5\grad$. As expected for $|B|>B_{crit}$ the two solutions in the GFT are nearly
the same and agree well with QMC. Below the critical field only the solution with the
slightly tilted field yields a stable homogeneous magnetization and its
$z$-component compares well with the QMC result in the untilted case.\\
\indent The above results can be interpreted within a semi-classical picture of non-collinear vortex
configurations which are stable below a critical field $B_{crit}$ in $z$-direction
and contribute a finite $z$-component to the magnetization in case of an applied field. 
\cite{ivanov02} Despite the lack of direct, quantitative access to such states 
(or corresponding physical in-plane observables) within the QMC algorithm they are included in 
principle and one can observe their consequences, 
namely a finite $z$-component of the magnetization below the critical GFT
field. On the other hand GFT can only describe homogeneous collinear 
configurations of spins therefore showing a breakdown of magnetization. However by applying
a small field in $x$-direction the $xy$-symmetry is broken and the spins rotate in the field
direction (the vortices vanish) and the collinear phase is retrieved.
Our results corroborate this interpretation based on the classical picture. Let us emphasize 
that both, GFT for slightly tilted field and QMC for $B\;||\;z$, describe the competition 
between the external field (which favors magnetization parallel to $z$) and the anisotropy favoring
in-plane magnetization. Comparing
the $z$-components of the magnetization for both cases, one can conclude that the ratio of the
competing forces are comparable for QMC and GFT. This indicates that this competition is 
correctly taken into account in GFT.\\
\indent In Fig. \ref{figure6} the same field
dependence of the $z$-component of the magnetization is shown for different temperatures. 
We have plotted the result for the tilted field in case of GFT, the point of breakdown in the 
untilted case is indicated by the dotted line. It can be seen that for higher temperatures no
breakdown of collinear magnetization occurs, meaning that the condition for the critical
field ($B \le |K_{eff} (T,B)|$) is never fulfilled in this case. The discrepancies at intermediate 
temperatures ($T=0.9..1.2$) are due to the RPA decoupling in the GFT as was discussed already.
\begin{figure}
\resizebox{0.95\columnwidth}{!}{
\includegraphics{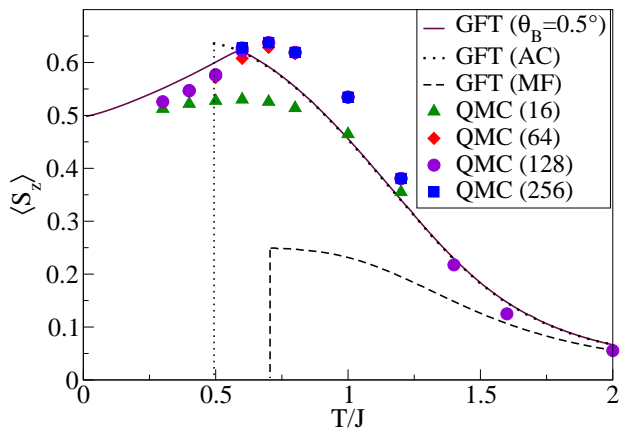}
}
\caption{The $z$-component of magnetization as function of temperature for a fixed
external field. Below a critical temperature $T_{crit}$ there is a breakdown of 
magnetization in GFT where is no in QMC. 
Parameters: $B/J=0.03$, $S=1$, $K_2/J=-0.06$.}\label{figure7}
\end{figure}

\begin{figure}
\resizebox{0.95\columnwidth}{!}{
\includegraphics{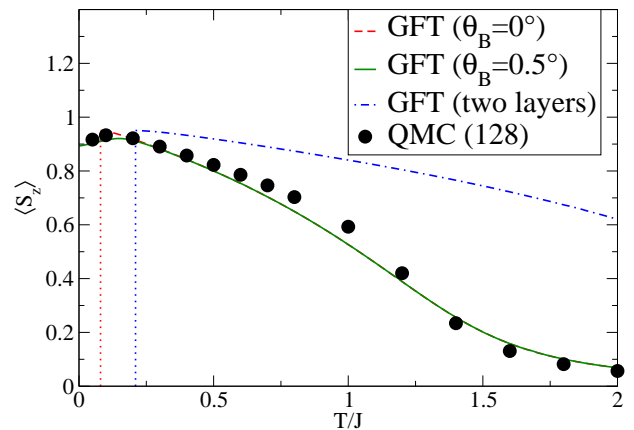}
}
\caption{Same situation as in Fig.~\ref{figure7} for $K_2/J=-0.04$ (other parameters as in
Fig.\ref{figure7}). The 
result for a two layer film treated by GFT is plotted also (dashed-dotted line).}\label{figure8}
\end{figure}
\begin{figure}
\resizebox{0.95\columnwidth}{!}{
\includegraphics{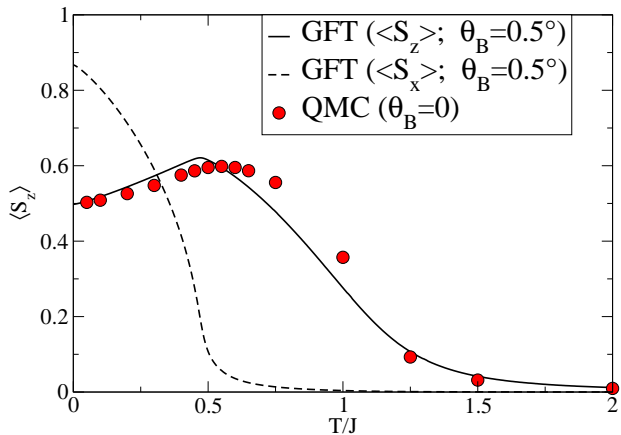}
}
\caption{Same situation as in Fig. \ref{figure7} for $K_2/J=-0.01$, $B/J=0.005$ and
slightly tilted field ($\theta_B=0.5\grad$) for the GFT results.}
\label{figure9}
\end{figure}

In Figs.~\ref{figure7}, \ref{figure8} and \ref{figure9} the $z$-component of the 
magnetization is plotted as a function of temperature obtained by GFT (straight line RPA+AC) 
as well as QMC (symbols) for different system sizes and a constant applied magnetic field.\\
\indent Let us first discuss the qualitative behavior of the magnetization as a function of 
temperature which is found in all three figures.
For high $T$ ($T\gg T_{crit}$) the magnetization is reduced by thermal fluctuations 
(where the tail of the curve above $T/J\approx 1.5$ is due to the
applied external field). In the vicinity of 
$T_{crit}$, $T-T_{crit}\rightarrow 0^+$, a competition between two effects sets in and has a 
pronounced influence on the magnetization.
On the one side the effective anisotropy acts against the external field 
($B_{eff}= B - |K_{eff}(T)|$, ($B\:||\:z$)). The effective anisotropy $K_{eff}(T)$ is reduced with
increasing temperature $T$ and thus the effective field $B_{eff}$ increases with $T$. 
This effect tends to enhance the magnetization with $T$. On the other side thermal 
fluctuations suppress the magnetization with
increasing $T$. The flattening of the magnetization curve near $T_{crit}$ is a result of 
this competition.
For low temperatures $T<T_{crit}$ the effective anisotropy in the GFT 
cannot be overcome by the external field ($B<|K_{eff}(T)|$, ($B\:||\:z$)).
Therefore the collinear magnetization in our approximation vanishes due to the
mentioned gapless excitations, in contrast to QMC which yields again a 
finite magnetization because non-collinear states are taken into account as discussed above.
The reduction of the $z$-component of magnetization in QMC 
below $T_{crit}$ can be pictured classically as the spins being in a non-collinear 
phase with an angle $\theta$ with respect to the $z$-axis. Since in general anisotropy 
effects (which favor in-plane magnetization) increase when temperature is lowered the 
$z$-component of the magnetization decreases. 

Now we discuss the three figures in detail. In Fig.~\ref{figure7} we have plotted QMC results
for different system size showing again that these are well converged for $N \geq 64$.
Thus we conclude that the striking difference between GFT and QMC is not a mere finite
size effect.
The breakdown of magnetization in GFT occurs at a critical temperature $T_{crit}/J=0.5$
whereas no such breakdown exists in QMC. However the exposed maximum of the magnetization
in QMC lies near the breakdown point. The differences between QMC and GFT in the temperature
range $T/J\approx0.3\dots1.3$ are due to the decoupling of the
exchange and anisotropy term in GFT as also seen in Fig.~\ref{figure3}.
It is worth mentioning that the value of the
z-component of the magnetization is nearly the same at the breakdown point in GFT and 
the maximum in the QMC. Thus we have the result that although GFT cannot describe 
the non-collinear phase by ansatz its breakdown coincides rather well with the onset 
of this phase, which we attribute to the maximum of the QMC curve. Fig.\ref{figure8}
shows the same situation for a different anisotropy constant $K_2=-0.04$. The critical 
temperature is lower than in Fig.\ref{figure7} since the ratio $B_z/K_2$ becomes larger. 
The tilted field case is also shown for the GFT results.
Again the qualitative agreement of the $z$-component of magnetization with QMC is good.
To confirm this point we have plotted the temperature dependence for an other set of 
parameters in Fig.~\ref{figure9}. There is as good qualitative agreement of the two approaches. 
Additionally one gets a finite component in $x$-direction in GFT which is also
plotted in the figure. The two effects of the external field vs. anisotropy competition
are nicely to be seen: a non-collinear state for $B\;||\;z$ ($z$-component only in QMC but not
in GFT) and rotation of magnetization for slightly tilted external field (seen only in GFT).
The ratio of the competing forces agree well again in both treatments.  

In Fig.~\ref{figure7} we have plotted the results of 
a different decoupling scheme of the anisotropy
terms (namely a mean field decoupling, dashed line in Fig.~\ref{figure7}).
Although the overall characteristic resembles the RPA$+$AC result (breakdown of magnetization)
the mean field results differ extremely from the QMC for a
large range of temperature and underestimates the magnetization.
This demonstrates the reliability of the Anderson-Callen treatment of the local
anisotropy terms presented in Refs.~\onlinecite{schwieger1,schwieger2,dipolpaper,pini}.

The extension of the GFT method to multi-layer films is straightforward.\cite{schwieger2} We 
have also included results for a two-layer film in Fig.~\ref{figure8} for the same 
parameters as in  the monolayer case. One finds that for a double layer magnetism is 
stabilized, which can be attributed to the increased coordination number and thus higher 
exchange energy. Just like for a monolayer, one observes a breakdown of collinear 
magnetization at some critical temperature. This is due to the fact that the same reasoning
regarding the vanishing excitation gap also applies for multilayer (slab) systems\cite{axel00}.
The effective anisotropy {\it per layer} 
is essentially the same as for a single layer, thus the critical $\langle S_{z} \rangle$-value 
(magnetization at critical field $B_{crit}$) is practically the same. The critical temperature 
is higher than that of a monolayer due to the increased magnetic stiffness of the double layer.

\section{Summary and Conclusions}

Using GFT and QMC calculations we studied easy-plane systems as well as easy-axis systems 
with an external field applied 
perpendicularly to the film. The GFT treatment of the Hamiltonian Eq.~(\ref{hamilt}) 
consists of a RPA-decoupling for the nonlocal terms
and an AC-decoupling for the local terms performed in a rotated frame, where the new $z'$-axis
is parallel to the magnetization. For the QMC calculations we have used the stochastic series
expansion (SSE) with directed loop updates, which is well suited for spin-systems.\\
\indent We have calculated the magnetization as a function of the external field as well as 
temperature. We found a critical field and critical temperature respectively below which is no 
magnetization in GFT whereas there is one in QMC. By tilting the field slightly in GFT so that 
it has a small component in $x$-direction we get a stable magnetization even below the critical
field or temperature. The $z$-component of the magnetization in this case coincides well with the
$z$-component obtained by QMC for the untilted field confirming that GFT and QMC agree well 
in the description of the external field vs. anisotropy competition. 
However, this comparison can be only somewhat indirect, since QMC has access to
the non-collinear ($B\;||\;z$) state only, while GFT is limited to collinear ferromagnetic
states (rotated homogeneous magnetization) found for slightly tilted external fields.\\ 
\indent For parameters that are accessible by both QMC and GFT 
($B\;||\;z$; $B>B_{crit}(T)$) QMC and GFT are in good agreement. Thus one can conclude that
the GFT is applicable to the homogeneous phases of systems described by Eq.~(\ref{hamilt})
and can be used also for system configurations 
not accessible by QMC due to too large system size as e.g. multilayer systems.\\
\indent It would be an interesting task for a succeeding work to extend the GFT
in order to get a deeper insight into the non-collinear configurations also.   
\newpage

\begin{appendix}
\section{magnetization angle} 
Here we will discuss the second mathematical solution which occurs besides Eq. 
\ref{2winkelloesungen}.
For an external field in the $z$-direction the angle dependent part of the free energy 
including second order anisotropy can be expanded  as \cite{lindner,farle}:

\begin{eqnarray*}
F = -M_zB_z\cos\theta-\tilde{K}_{2}\cos^2\theta
\end{eqnarray*}

where $M_z$ is the $z$-component of the magnetization and $\tilde{K}_{2}$ is the first
nonvanishing coefficient in an expansion of the free energy for a system with second 
order anisotropy.
For the equilibrium angle one gets:

\begin{eqnarray}
\frac{\partial F(\theta)}{\partial\theta} &=& 
M_zB_z\sin\theta+2\tilde{K}_{2}\cos\theta\sin\theta \stackrel{!}{=} 0.
\end{eqnarray}
Therefore one gets two solutions for in-plane systems ($\tilde{K}_{2}<0$). 
For $\sin\theta\neq 0$ one gets immediately the solution of Eq. \ref{2winkelloesungen} if 
$2\tilde{K}_{2}/M_z\equiv K_{eff}$ holds. This is the stable solution.
The trivial (second) solution $\sin\theta=0$ is unstable for $B_z<|K_{eff}|$ because
\begin{eqnarray}
\frac{\partial^2 F(\theta)}{\partial\theta^2}\mid_{\sin\theta=0} &=& 
\left\{ \begin{array}{cc} <0 & ~~\text{for}~~B_z<|K_{eff}|\\
 >0 & \text{otherwise} 
 \end{array}\right.
\end{eqnarray}
holds. For a detailed discussion of stability conditions in film systems we refer to 
Refs.~\onlinecite{farle,lindner}.

\end{appendix}


\begin{thebibliography}{}

\bibitem{farle} M.~Farle, B.~Mirwald-Schulz, A.~N.~Anisimov, W.~Platow, and 
K.~Baberschke, 
Phys. Rev. B \textbf{55}, 3708 (1997).

\bibitem{wandlitz17} A.~Hucht and K.~D.~Usadel, 
Phys. Rev. B \textbf{55}, 12309 (1997).

\bibitem{wandlitz18} P.~J.~Jensen and K.~H.~Bennemann, 
Solid State Comm. \textbf{105}, 577 (1998), and references therein. 

\bibitem{wandlitz19} R.~P.~Erickson and D.~L.~Mills, 
Phys. Rev. B \textbf{44}, 11825 (1991). 

\bibitem{wandlitz20} D.~K.~Morr, P.~J.~Jensen, and K.~H.~Bennemann, 
Surf. Sci. \textbf{307-309}, 1109 (1994). 

\bibitem{wandlitz21} P.~Politi, A.~Rettori, M.~G.~Pini, and D.~Pescia, 
J. Magn. Magn. Mater. \textbf{140-144}, 647 (1995);\\ 
A.~Abanov, V.~Kalatsky, V.~L.~Pokrovsky and W.~M.~Saslow, 
Phys. Rev. B \textbf{51}, 1023 (1995). 

\bibitem{wandlitz22} A.~Hucht, A.~Moschel, and K.~D.~Usadel, 
J. Magn. Magn. Mater. \textbf{148}, 32 (1995);\\
S.~T.~Chui, Phys. Rev. B \textbf{50}, 12559 (1994). 

\bibitem{wandlitz24} T.~Herrmann, M.~Potthoff, and W.~Nolting, 
Phys. Rev. B \textbf{58}, 831 (1998).

\bibitem{qm} P.~Henelius, P.~Fr\"obrich, P.~J.~Kuntz, C.~Timm, and 
P.~J.~Jensen, Phys. Rev. B \textbf{66}, 094407 (2002).

\bibitem{schwieger1} S.~Schwieger, J.~Kienert, and W.~Nolting, 
Phys. Rev. B \textbf{71}, 024428 (2005). 

\bibitem{schwieger2} S.~Schwieger, J.~Kienert, and W.~Nolting, 
Phys. Rev. B \textbf{71}, 174441 (2005).

\bibitem{dipolpaper} F.~K\"ormann, S.~Schwieger, J.~Kienert, and W.~Nolting, 
Eur. Phys. J. B \textbf{53}, 463 (2006).

\bibitem{pini} M.~G.~Pini, P.~Politi and R.~L.~Stamps, 
Phys. Rev. B \textbf{72}, 014454 (2005).

\bibitem{kosterlitz73} J.~M.~Kosterlitz and D.~J.~Thouless, 
J. Phys. C \textbf{6}, 1181 (1973).

\bibitem{wysin98} G.~M.~Wysin, 
Phys. Lett. A \textbf{240}, 95 (1998).

\bibitem{vedmedenko99} E.~Yu.~Vedmedenko, A.~Ghazali, and J.~-C.~S.~L\'evy, 
Phys. Rev. B \textbf{59}, 3329 (1999).

\bibitem{lee04} K.~W.~Lee and C.~E.~Lee, 
Phys. Rev. B \textbf{70}, 144420 (2004).

\bibitem{rapini07} M.~Rapini, R.~A.~Dias, and B.~V.~Costa, 
Phys. Rev. B \textbf{75}, 014425 (2007).

\bibitem{ivanov02} B.~A.~Ivanov and G.~M.~Wysin, 
Phys. Rev. B \textbf{65}, 134434 (2002).

\bibitem{bloch30} F.~Bloch, Z. Phys. \textbf{61}, 206 (1930).

\bibitem{bruno91} P.~Bruno, Phys. Rev. B \textbf{43}, 6015 (1998).

\bibitem{wandlitz} P.~J.~Jensen and K.~H.~Bennemann, in \textit{Magnetism and 
electronic correlations in local-moment systems}, edited by M.~Donath, 
P.~A.~Dowben and W.~Nolting, p.113 (World Scientific, 1998).

\bibitem{jensen2000} P.~Fr\"obrich, P.~J.~Jensen, and P.~J.~Kuntz, 
Eur. Phys. J. B \textbf{13}, 477 (2000).

\bibitem{RPA} N.~N.~Bogolyubov and S.~V.~Tyablikov, 
Soviet. Phys.-Doklady \textbf{4}, 589 (1959).

\bibitem{ac} F.~B.~Anderson and H.~Callen, 
Phys. Rev. \textbf{136}, A1068 (1964).

\bibitem{exakt} P.~Fr\"obrich and P.~J.~Kuntz, http://arxiv.org/pdf/cond-mat/0607675.

\bibitem{lindner} J.~Lindner, 
\textit{Ph.D. thesis}, Freie Universit\"at Berlin (2002).

\bibitem{sandvik99} A.~W.~Sandvik, 
Phys. Rev. B \textbf{59}, R14 157 (1999).

\bibitem{sandvik02} O.~F.~Sylju\aa{}sen and A.~W.~Sandvik, 
Phys. Rev. E \textbf{66}, 046701 (2002).

\bibitem{alet05} F.~Alet, S.~Wessel, and M.~Troyer, 
Phys. Rev. E \textbf{71}, 036706 (2005).

\bibitem{ALPS} ALPS collaboration, 
J. Phys. Soc. Jpn. Suppl. \textbf{74}, 30 (2005). 
Source codes can be obtained from http://alps.comp-phys.org/

\bibitem{axel00} A.Gelfert and W.Nolting,
Phys. Stat. Sol. B \textbf{217}, 805 (2000).
\end{thebibliography}
\end{document}